\documentclass[preprint,aps,showpacs,preprintnumbers,amsmath,amssymb,nofootinbib]
{revtex4}

\usepackage{epsfig}
\begin{document}

\begin{flushright}
\end{flushright}


\newcommand{\be}{\begin{equation}}
\newcommand{\ee}{\end{equation}}
\newcommand{\bea}{\begin{eqnarray}}
\newcommand{\eea}{\end{eqnarray}}
\newcommand{\nn}{\nonumber}
\def\CP{{\it CP}~}
\def\cp{{\it CP}}
\title{\large Charged lepton correction to tribimaximal lepton mixing  and its implications to neutrino phenomenology }
\author{Srinu Gollu, K.N. Deepthi, R. Mohanta }
\affiliation{
School of Physics, University of Hyderabad, Hyderabad - 500 046, India }

\begin{abstract}

The  recent results from Daya Bay and RENO reactor neutrino experiments have firmly established that the smallest reactor mixing angle $\theta_{13}$ is
non-vanishing at the $5 \sigma$ level, with a relatively large value, i.e., $\theta_{13}\approx 9^{\circ}$. Using the fact that the neutrino
mixing matrix can be represented as  $V_{\rm PMNS}=U_l^{\dagger} U_{\nu} P_\nu$, where $U_l$ and $U_\nu$ result from the diagonalization
of the charged lepton and  neutrino mass matrices and $P_\nu$ is a diagonal matrix containing the Majorana phases and
assuming the tri-bimaximal form for $U_\nu$, we investigate the possibility
of accounting for the large reactor mixing angle due to the corrections of the charged lepton mixing matrix.
The form of $U_{l}$ is assumed to be that of CKM mixing matrix of the quark sector.
We find that with this modification it is possible to accommodate the large observed reactor mixing angle $\theta_{13}$. We also study the implications of such corrections on the other phenomenological observables.\\

\end{abstract}

\maketitle

\section{Introduction}

The results from various neutrino oscillation experiments firmly established the fact that
neutrinos have a tiny but finite nonzero mass. Thus,
analogous to the mixing in the down-quark sector, the  three
flavor eigenstates of neutrinos ($\nu_e,~ \nu_\mu,~ \nu_\tau$)
are related to the corresponding
mass eigenstates ($\nu_1,~ \nu_2,~ \nu_3$) by the unitary transformation
\bea
\left( \begin{array}{c}
 \nu_e       \\
\nu_\mu \\
\nu_\tau \\
\end{array}
\right ) \; =\; \left ( \begin{array}{ccc}
V_{e1}      & V_{e2}    & V_{e3} \\
V_{\mu 1}      & V_{\mu 2}    & V_{\mu 3} \\
V_{\tau 1}      & V_{\tau 2}    & V_{\tau 3} \\
\end{array}
\right ) \left ( \begin{array}{c}
\nu_1 \\
\nu_2 \\
\nu_3 \\
\end{array}
\right ) \; ,
\eea
where $V$ is the $3 \times 3 $ unitary
matrix known as PMNS matrix \cite{pmns}, which contains three mixing angles
and three CP violating phases (one Dirac type and two Majorana
type).  In the standard
parametrization \cite{std}, $V_{\rm PMNS}$ is expressed
in terms of the solar, atmospheric and reactor mixing angles
 $\theta_{12}^{}$,
$\theta_{23}^{}$, $\theta_{13}^{}$ and three CP-violating phases
$\delta_{\rm CP}, \rho, \sigma$ as
\begin{eqnarray}
V_{\rm PMNS} = \left( \begin{array}{ccc} c^{}_{12} c^{}_{13} & s^{}_{12}
c^{}_{13} & s^{}_{13} e^{-i\delta_{\rm CP}} \\ -s^{}_{12} c^{}_{23} -
c^{}_{12} s^{}_{13} s^{}_{23} e^{i\delta_{\rm CP}} & c^{}_{12} c^{}_{23} -
s^{}_{12} s^{}_{13} s^{}_{23} e^{i\delta_{\rm CP}} & c^{}_{13} s^{}_{23} \\
s^{}_{12} s^{}_{23} - c^{}_{12} s^{}_{13} c^{}_{23} e^{i\delta_{\rm CP}} &
-c^{}_{12} s^{}_{23} - s^{}_{12} s^{}_{13} c^{}_{23} e^{i\delta_{\rm CP}} &
c^{}_{13} c^{}_{23} \end{array} \right) P^{}_\nu \equiv U_{\rm PMNS} P_\nu\;
,\label{mat}
\end{eqnarray}
where $c^{}_{ij}\equiv \cos \theta^{}_{ij}$, $s^{}_{ij} \equiv \sin
\theta^{}_{ij}$ and $P_\nu^{} \equiv \{ e^{i\rho}, e^{i\sigma}, 1\}$ is
a diagonal matrix with CP violating Majorana phases $\rho$ and $\sigma$.
The $U_{\rm PMNS}$ component of the mixing matrix describes the mixing of Dirac type
neutrinos analogous to the CKM matrix of the quark sector.
The neutrino oscillation data accumulated over many years allow us to determine the solar
and atmospheric neutrino oscillation parameters with very high precision.
Recently, the value of the smallest mixing angle $\theta_{13}$ has been measured by the Daya Bay \cite{dayabay} and RENO Collaborations
\cite{reno} with the
 best fit ($ 1 \sigma $) result as
\bea
\sin^2 2 \theta_{13}&=&0.089 \pm 0.010({\rm stat}) \pm 0.005({\rm syst}),~~~~{\rm Daya~ Bay}\nn\\
\sin^2 2 \theta_{13}&=&0.113 \pm 0.013({\rm stat}) \pm 0.019({\rm syst}).~~~~~{\rm RENO}
\eea
which is equivalent to $\theta_{13} \simeq 8.8^\circ \pm 0.8^\circ $. This is   $5.2 \sigma$
evidence of nonzero value of  $ \theta_{13}$ which confirms the previous measurements of T2K \cite{T2K}, MINOS
\cite{minos} and
Double Chooz \cite{chooz} experiments.
The global analysis of the recent results of various neutrino oscillation experiments has been performed by
several groups \cite{gfit1,gfit2,gfit3}, and the parameters which are used in this analysis are taken from Ref.
\cite{gfit3}, are presented in Table-1.

\begin{table}
\caption{The global fit values of the mixing parameters taken from \cite{gfit3}.}

\begin{tabular}{|c|c|c|}
\hline
 Mixing Parameters & $1 ~\sigma $ value & $ 3 ~\sigma $ Range  \\
\hline
$\sin^2 \theta_{12} $ &~ $0.302_{-0.012}^{+0.013}$ ~& ~$ 0.267 \to 0.344 $~\\

$\theta_{12}/ {^\circ}$ & ~ $33.36_{-0.78}^{+0.81} $~ & ~ $~31.09 \to 35.89$~\\

$\sin^2 \theta_{23} $ &~ $0.413_{-0.025}^{+0.037}$ ~& ~$ 0.342 \to 0.667 $~\\

$\theta_{23}/ {^\circ}$ & ~ $40.0_{-1.5}^{+2.1} $~ & ~ $~35.8 \to 54.8$~\\

$\sin^2 \theta_{13} $ &~ $0.0227_{-0.0024}^{+0.0023}$ ~& ~$ 0.0156 \to 0.0299 $~\\

$\theta_{13}/ {^\circ}$ & ~ $8.66_{-0.46}^{+0.44} $~ & ~ $~7.19 \to 9.96$~\\

$\delta_{\rm CP}/{^\circ}$ & ~$300_{-138}^{+66}~$ & $ ~0 \to 360~ $\\
$\Delta m_{21}^2/ 10^{-5} {\rm eV}^2 $ & $ 7.5_{-0.19}^{+0.18} $ & $ 7.00 \to 8.09 $ \\

$\Delta m_{31}^2/ 10^{-3} {\rm eV}^2 ({\rm NH}) $~ &~ $ 2.473
_{-0.067}^{+0.07} $ & $ 2.276 \to 2.695 $ \\

$\Delta m_{32}^2/ 10^{-3} {\rm eV}^2 ({\rm IH}) $ ~&~ $ -2.427_{-0.065}^{+0.042} $ & $ -2.649 \to -2.247 $ \\

\hline
\end{tabular}
\end{table}

The observation of this not so small reactor mixing angle $\theta_{13}$  has ignited a lot of interest to understand the mixing
pattern in the lepton sector \cite{xing}. It opens promising perspectives for the observation of CP violation in the
 lepton sector. The precise determination  of $\theta_{13}$ in addition to providing a complete picture of neutrino mixing pattern, could be a signal of underlying physics responsible for lepton mixing and for the physics beyond standard model.
It has been shown that if  one includes some perturbative corrections to the leading order neutrino
mixing patterns, such as bi-maximal (BM) \cite{bm}, tri-bimaximal (TBM) \cite{tbm} and democratic (DC) \cite{dc}, it is possible to explain the  observed neutrino mixing angles \cite{rm}.
However, it should be noted that among these leading order mixing patterns i.e., BM, TBM and DC, the
tri-bimaximal pattern, whose explicit form as  given below
\begin{eqnarray}
U_{\rm TBM} &=& \left ( \begin{array}{ccc}
~\sqrt{\frac{2}{ 3}}&~\sqrt{\frac{1}{3}}&~~0\\
-\sqrt{\frac{1}{6}}&~\sqrt{\frac{1}{ 3}}&~\sqrt{\frac{1}{2}}\\
-\sqrt{\frac{1}{ 6}}&~\sqrt{\frac{1}{ 3}}& -\sqrt{\frac{1}{2}}
\end{array}
\right )\;,\label{tbm}
\end{eqnarray}
is particularly very interesting. It corresponds to the three mixing angles of the standard parametrization as
 $\theta_{12}=\arctan(1/\sqrt 2) \simeq35.3^\circ$, $\theta_{13}=0^\circ$ and $\theta_{23}=45^\circ$.
Clearly, to accommodate the large value of $\theta_{13}$, one has to consider possible perturbations to the
TBM mixing matrix. In this paper we would like to study the possible corrections arising from the charged
lepton sector. The essential features of our analysis are as follows. We assume the charged lepton mixing matrix to be
of the same form as the CKM quark mixing  matrix and the neutrino mixing matrix to be of the tri-bimaximal form.
Furthermore, we use the Wolfenstein-like parametrization for the charged lepton mixing matrix and study its implications
on various phenomenological observables. It should be noted that there have been several attempts made recently
to understand the nonzero $\theta_{13}$ due to
charged lepton correction \cite{lepton} and  in the past also  corrections to the leptonic mixing matrix due
to charged leptons were considered in Ref. \cite{lepton-1}.

The paper has been organized as follows. The methodology of our analysis is presented in Section-II and the
Results and Conclusion are discussed in Section-III.

\section{Methodology}

It is well known that the leptonic mixing matrix arises from the overlapping of the matrices that diagonalize
charged lepton and neutrino mass matrices
\bea
U_{\rm PMNS} = U_l^\dagger U_\nu \label{mix}\;.\eea
Here we are focussing only the component of the mixing matrix which describes the mixing of Dirac type
neutrinos.
For the study of leptonic mixing  it is generally assumed that the charged lepton mixing matrix as an identity
matrix and the neutrino mixing matrix $U_\nu$ has a specific form dictated by a symmetry
 which fixes the values of the three mixing angles in $U_\nu$. The small deviations of
the mixing angles from those measured in the PMNS matrix, are considered, in general, as  perturbative corrections
arising from symmetry breaking effects.  A variety  symmetry forms of $U_\nu$ have been explored in the literature e.g.,
BM/TBM/DC and so on. In this work we will consider the situation wherein the neutrino mixing matrix is described by the
TBM matrix, i.e.,
\be
 U_{\nu}= U_{\rm TBM}\;, \label{nu}
\ee
 and that
the mixing angles induced by the charged leptons can be considered as corrections.
Furthermore, we will neglect possible corrections to $U_{\rm TBM}$ from higher dimensional operators and from renormalization group effects.
In this approximation we will derive formulae which allow us to include corrections to neutrino mixing angles and
to constrain the CP violating phase ($\delta_{CP}$) conveniently.

In our study, we use a simple {\it ansatz} for the charged lepton mixing matrix $U_l$, i.e., we assume that $U_l$ has the same structure as the CKM matrix connecting the  weak eigenstates of the down type quarks to the corresponding mass eigenstates.  This approximation is
quite reasonable as we know that the CKM matrix is almost diagonal with the off diagonal elements strongly suppressed by the
small expansion parameter $\lambda = \sin \theta_C $ ($\theta_C$, being the Cabibbo angle).
Hence, such an assumption can naturally provide
the small perturbations to the tri-bimaximal mixing pattern for neutrino mixing matrix.
Furthermore, as discussed in Ref. \cite{cheng}, this approximation is quite acceptable as the mass spectrum of charged leptons exhibits similar hierarchical structure
as the down type quarks, i.e., $(m_e,m_\mu) \approx (\lambda^5, \lambda^2) m_\tau$ and $(m_d, m_s) \approx (\lambda^4, \lambda^2) m_b$. This may imply that the charged lepton mixing matrix has a structure similar to the down type quark mixing and
is governed by the CKM matrix.

To illustrate the things more explicitly, let us recall the values of the quark mixing angles in the standard PDG parametrization for the CKM matrix within $1\sigma$ range as \cite{PDG}
\be
\theta_{13}^{q} = 0.20^\circ \pm 0.01^\circ,~~~~\theta_{23}^{q} = 2.35^\circ \pm 0.07^\circ, ~~~~
\theta_{12}^{q}\equiv \theta_C =13.02^\circ \pm 0.04^\circ.
\ee
However, the leptonic sector is described by two large mixing angles  $\theta_{23}^{l}$ and $\theta_{12}^{l}$ and the  third mixing angle
$\theta_{13}^l$, was expected to be very small. Recently, the third mixing angle $\theta_{13}^l$  has been measured by T2K, Double CHOOZ, Daya Bay and RENO
Collaborations yielding the following mixing patterns in the lepton sector:
\be
\theta_{13}^{l}= 8.8  ^\circ \pm 1.0^\circ,~~~~~\theta_{23}^{l}= 40.4^\circ \pm 1.0^\circ,~~~~
\theta_{12}^{l}= 34.0^\circ \pm 1.1^\circ\;.
\ee
The different nature of the quark and lepton mixing angles can be inter-related in terms of the quark lepton
complementarity (QLC) relations \cite{QLC}, as
\be
\theta_{12}^{q} + \theta_{12}^{l} \simeq 45^\circ,~~~~~~~~\theta_{23}^{q} + \theta_{23}^{l} \simeq 45^\circ.
\ee
The QLC relations indicate that it could be possible to have a quark-lepton symmetry based on some flavor symmetry.
The experimental result of this not-so-small reactor mixing angle $\theta_{13}^l$ has triggered a lot of interest
in the theoretical community.
Given the rather precise measurement of $\theta_{13}^{l}$, one may wonder  whether
$\theta_{13}^{l}$ numerically agrees well with the QLC relation, i.e.,
\be
\theta_{13}^{l} = \frac{\theta_C}{\sqrt 2} \approx 9.2^\circ.
\ee
In particular, it is quite interesting to see  whether this specific connection to $\theta_C$ can be a consequence of some underlying symmetry,
which  may provide a clue to the nature of quark-lepton physics beyond the standard model.

Starting from the fact that the mixing matrix of the up type quark sector can be almost diagonal and so the CKM matrix
is mainly generated from the down type quark mixing matrix, we assume that the mixing matrix of the charged
lepton sector is basically of the same form as that of down type quark sector.
 Consequently the lepton mixing matrix appears as the product of CKM like matrix (induced by charged lepton sector)
and the TBM pattern matrix induced from the neutrino sector. As discussed before, in the limit of diagonal charged lepton mass matrix i.e., $U_l ={\bf 1}$, and $U_{\nu}=U_{\rm TBM}$, which gives the
mixing angles at the leading order as
\bea
\theta_{12}^{l 0}=\arctan(1/\sqrt 2) \simeq 35.3^\circ,~~ \theta_{13}^{l0}=0^\circ,~~~{\rm  and}~~~~~ \theta_{23}^{l0}=45^\circ\;,
\eea
 deviate
significantly from their measured values as
\be
|\theta_{12}^l -\theta_{12}^{l0}|  \simeq 2^\circ,~~~\theta_{13}^l-\theta_{13}^{l0} \approx 9^\circ~~~~ {\rm and} ~~~~
|\theta_{23}^l-\theta_{23}^{l0}| \simeq 5^\circ\;.
\ee
These deviations are attributed to the corrections arising from the charged lepton sector.
We assume the charged lepton mixing matrix to have the form as the CKM matrix in the standard parametrization, i.e.,
\be
U_l = R_{23} U_{13} R_{12}\;,
\ee
where the matrices $R_{23}$, $U_{13}$ and $R_{12}$ are defined by
\begin{eqnarray}
R_{12} &=& \left (\begin{array}{ccc}
\cos \theta_{12}^l & \sin \theta_{12}^l &0\\
-\sin \theta_{12}^l &\cos \theta_{12}^l &0\\
0&0&1
\end{array}
\right )\;, \;\;~~~~~~~~~~~~~ R_{23} = \left (\begin{array}{ccc}
1&0&0\\
0&\cos \theta_{23}^l  &\sin \theta_{23}^l\\
0&-\sin \theta_{23}^l& \cos \theta_{23}^l
\end{array}\right )\nn\\
U_{13} & = & \left ( \begin{array}{ccc}
\cos \theta_{13}^l &0&\sin \theta_{13}^l ~ e^{-i\delta} \\
0&1&0\\
-\sin \theta_{13}^l ~e^{i \delta} &0& \cos \theta_{13}^l
\end{array}\right )~.
\label{vb}
\end{eqnarray}
Furthermore, as
 the mixing angle  $\theta_{13}$ receives maximum deviation from the TBM pattern, we assume
 $\sin \theta_{13}^l= \sin \theta_C = \lambda$, where, $\lambda$ is a small expansion parameter
 analogous to the expansion parameter of Wolfenstein parametrization of the CKM matrix.
 The other two angles are assumed to be of the form
 \be
 \sin \theta_{23}^l= A \lambda^2,~~~~~~~\sin \theta_{12}^l = A \lambda^3\;,
 \ee
 where the parameter $A={\cal O}(1)$.
With these values, one can obtain the  Wolfenstein-like parametrization for $U_l$ (upto order $\lambda^3$) as
\begin{eqnarray}
U_{l} & = & \left ( \begin{array}{ccc}
1- \lambda^2/2 &  A \lambda^3  &\lambda ~ e^{-i\delta} \\
-A\lambda^3(1+e^{i\delta}) &1&A \lambda^2\\
-\lambda e^{i \delta} & -A \lambda^2 & 1-\lambda^2/2
\end{array}\right )
\label{Ul}
\end{eqnarray}

Thus, with the help of Eqs.~(\ref{mix}), (\ref{nu}) and (\ref{Ul}), one can schematically obtain the PMNS matrix
up to order of $\lambda^{3}$ as
\be
 U_{\rm PMNS}=U_{\rm TBM}+\Delta U\;,\label{dphase}
\ee
with
 \begin{eqnarray}
\Delta  U = {\left(\begin{array}{ccc}
  \frac{ \lambda e^{-i\delta} -\lambda^2 +A \lambda^3 (1+e^{-i \delta})}{\sqrt 6} &
 -\frac{ \lambda e^{-i\delta} +\lambda^2/{2} +A \lambda^3 (1+e^{-i \delta}) }{\sqrt 3}
 &
 -\frac{ \lambda e^{-i\delta}-A \lambda^3 (1+e^{-i \delta})}{\sqrt 2}
\\
 \frac{A \lambda^2(1+2\lambda)}{\sqrt 6} & -\frac{A \lambda^2(1-\lambda)}{\sqrt 3} &
-\frac{A \lambda^2}{\sqrt{2}} \\
   \frac{2\lambda e^{i \delta} -A \lambda^2 + \lambda^2/2}{\sqrt 6}  &
\frac{\lambda e^{i \delta} + A \lambda^2 - \lambda^2/2}{\sqrt 3} &
-\frac{A\lambda^2+\lambda^2/2}{\sqrt{2}} \\
   \end{array}\right)}
   +{\cal O}(\lambda^4)\;,
 \label{PMNS}
 \end{eqnarray}
which  allows one to obtain the elements of the PMNS matrix  as
\begin{eqnarray}
  |U_{e1}|&=& \sqrt{\frac{2}{3}}\left [1+\frac{1}{2}\lambda\cos\delta-\frac{1}{8}\lambda^{2}
  (3+\cos^{2}\delta)+\frac{1}{16}\lambda^{3} \left (8 A(1 +\cos\delta )-\cos\delta \sin^2 \delta \right )\right ],\nonumber\\
  |U_{e2}|&=& \frac{1}{\sqrt{3}}\left[1-\lambda\cos\delta-\frac{1}{2}\lambda^{2}\cos^{2}
  \delta - \frac{1}{2}\lambda^{3}(2A(1+\cos \delta)-\cos\delta \sin^2 \delta)\right]~,\nonumber\\
  |U_{e3}|&=& \frac{\lambda}{\sqrt{2}}\left [1-A \lambda^2(1+\cos \delta)\right ]~,\nonumber\\
  U_{\mu1}&=& -\frac{1}{\sqrt{6}}\left[1 -A \lambda^2 -2 A \lambda^3 \right]~,\nonumber\\
  U_{\mu2}&=& \frac{1}{\sqrt{3}}\left[1- A\lambda^2+ A \lambda^3 \right]~,\nonumber\\
  |U_{\mu3}| &=& \frac{1}{\sqrt{2}}\left(1+ A\lambda^{2}\right)~,\nonumber\\
  U_{\tau1} &= &- \frac{1}{\sqrt{6}}\left(1-2 \lambda e^{i \delta} +\frac{1}{2}\lambda^{2}(2 A - 1) \right)~,\nonumber\\
  U_{\tau2}&=& \frac{1}{\sqrt{3}}\left(1+\lambda e^{i \delta} + \frac{1}{2} \lambda^{2}(2 A-1)\right)~,\nonumber\\
  |U_{\tau3}| &= & \frac{1}{\sqrt{2}}\left(1- \frac{1}{2} \lambda^{2}- A \lambda^2)\right)~.
  \label{elements1}
 \end{eqnarray}
From Eq.~(\ref{mat}), one can express  the neutrino mixing parameters in terms of the PMNS mixing matrix elements  as
  \begin{eqnarray}
  \sin^{2}\theta_{12}&=&\frac{|U_{e2}|^{2}}{1-|U_{e3}|^{2}}~,\qquad\qquad\quad
   \sin^{2}\theta_{23}=\frac{|U_{\mu3}|^{2}}{1-|U_{e3}|^{2}}~,\nonumber\\
  \sin\theta_{13}&=&|U_{e3}|~.
 \label{mixing1}
 \end{eqnarray}
Thus, from Eqs. (\ref{elements1}) and (\ref{mixing1}), one can obtain the  solar neutrino mixing angle $\theta_{12}$, up to order $\lambda^3$, as
 \begin{eqnarray}
  \sin^{2}\theta_{12}\simeq\frac{1}{3}\left(1-2\lambda\cos\delta+\frac{\lambda^2}{2}-\lambda^3
[2A(1 + \cos \delta) + \cos^3 \delta] \right),
 \label{theta12}
 \end{eqnarray}
Clearly, when $ \cos \delta$ approaches  zero we observe a tiny deviation from $\sin^2\theta_{12}= { 1/3}$.
Following  similar approach, one can obtain the atmospheric neutrino mixing angle $\theta_{23}$ as
 \begin{eqnarray}
   \sin^{2}\theta_{23}\simeq\frac{1}{2}\left(1+\frac{\lambda^{2}}{2}(1+4 A)\right)~,
 \label{theta23}
 \end{eqnarray}
which also shows a small deviation from the maximal mixing pattern i.e., $\sin^2\theta_{23}= { 1/2}$.
The reactor mixing angle $\theta_{13}$ can be obtained as
 \begin{eqnarray} \label{Diracphase}
  \sin\theta_{13}&=&\frac{\lambda}{\sqrt{2}}\left (1-A\lambda^{2}(1+\cos\delta)\right )~.
 \label{theta13}
 \end{eqnarray}
Thus, we have a non-vanishing large $\theta_{13}$. This in turn implies that it could be possible to observe
CP violation in the lepton sector analogous to the quark sector, which
could be detected through long base-line neutrino oscillation experiments. The Jarlskog invariant,
which is a measure of CP violation, for the lepton sector has the expression
 \begin{eqnarray}
  J^{\ell}_{\rm CP}\equiv{\rm Im}[U_{e 1}U_{\mu 2}U^{\ast}_{\mu 1}U^{\ast}_{e 2 }]=-\frac{\lambda\sin\delta}{6}
  \left(1-\frac{\lambda^{2}}{2}- A \lambda^2) \right)+{\cal O}(\lambda^4)~,
 \label{Jcp2}
 \end{eqnarray}
which is sensitive to the Dirac CP violating phase.

The Dirac CP phase $\delta_{CP}$ can be deduced by using the PMNS matrix elements and the neutrino mixing parameters as \cite{cheng}
\begin{eqnarray}
  \delta_{\rm CP}
=-\arg \left(\frac{\frac{U^{\ast}_{e1}U_{e3}U_{\tau1}U^{\ast}_{\tau3}}{c_{12}c^{2}_{13}c_{23}s_{13}}+c_{12}c_{23}s_{13}}{s_{12}s_{23}}\right)~.
\label{DCP}
\end{eqnarray}
With Eqs. (\ref{tbm}), (\ref{dphase}) and (\ref{PMNS}), this yields the correlation between the two CP violating phases (Dirac type CP violating phase and
the phase $\delta$ introduced in the charged lepton mixing)
\bea
\delta_{\rm CP} = - \arctan \left [\frac{- \lambda(1-(A+ \frac{1}{2} \lambda^2)) \sin \delta}{\lambda \Big[(A(1- \lambda^2) -\frac{5}{2} \lambda^2)\cos \delta
-(\frac{3}{2}\lambda+A \lambda^2)\Big]+ \lambda^2(1+ \lambda \cos \delta)} \right ]\;.
\eea

Three mass-dependent neutrino observables are probed in different types of experiments.
The sum of absolute neutrino masses $ \sum_i m_i$ is probed in cosmology, the kinetic electron neutrino mass 
in beta decay ($M_\beta$ ) is probed in direct search
for neutrino masses, and the effective mass ($M_{ee}$ ) is probed in neutrino-less double
beta decay experiments with the decay rate for the process $\Gamma \propto |M_{ee}|^2$. In terms of
the bare physical parameters $m_i$ and $U_{\alpha i}$, the observables are given by \cite{dbd}
\bea
&&\sum_i m_i  =  m_1  + m_2  + m_3 ,\nn\\
&&M_{ee} = \sum_i U_{ei} ^2  m_i ,\nn\\
&&M_{\beta} =   \sqrt{\sum_i |U_{ei} |^2 m_i ^2 }\;. \label{abs}
\eea
The absolute values of neutrino masses are currently unknown.
Recently the Planck experiment on Cosmic Microwave Background (CMB) \cite{cmb} has reported an interesting result 
for the sum of three neutrino masses  with an assumption of three species of degenerate neutrinos 
as
\be
\sum_i m_i \leq 0.23~{\rm eV}\;.~~~ ({\rm Planck+WP+highL+BAO})
\ee
The most stringent upper bound on electron-antineutrino mass has been measured 
in the  Troitsk experiment \cite{trot} on the high precision measurement of the end-point spectrum of 
tritium beta decay as 
\be
M_\beta < 2.05~{\rm eV}~~~95\% ~{\rm C.L.}
\ee

In our analysis we ignore the Majorana phases $(\rho , \sigma )$ and  consider the normal hierarchy scenario
for the neutrino mass spectrum in which the neutrino masses $m_2$ and $m_3$ can be expressed in terms of the
lightest neutrino mass $m_1$ and the measured solar and atmospheric mass-squared differences $\Delta m_{\rm sol}^2$ and
$\Delta m_{\rm atm}^2 $ as
\be
m_2= \sqrt{m_1^2 + \Delta m_{\rm sol}^2},~~~~~~m_3= \sqrt{m_1^2 + \Delta m_{\rm sol}^2 + \Delta m_{\rm atm}^2}\;.
\ee

\section{Results and Discussion}

For numerical estimation we need to know the values of the three unknown parameters $A$, $\lambda$ and $\delta$.
In this analysis we assume the small expansion parameter $\lambda$ to have the same value as that of the quark sector
\cite{PDG}:
\be
\lambda = 0.22535 \pm 0.00065\;.
\ee
Now with Eq. (\ref{theta23}) and using the experimental value of $\sin^2 \theta_{23}$ as input parameter, we obtain the  $1\sigma~(
3 \sigma) $ range of $A$ as
\bea
A&=&( -2.4 \to -1.2)~~~~(1\sigma)\nn\\
 &=&( -3.4 \to 3.0),~~~~(3\sigma)
\eea
and we treat the CP violating phase $\delta$ as a free parameter, i.e., we allow it to vary in its entire range
$0 \leq \delta \leq 2 \pi$.
Now varying these  input parameters in their $3\sigma$ ranges, and using Eqs. (\ref{theta12}) and (\ref{theta13}),
we present the variation of  the solar and reactor mixing angles ($\sin^2\theta_{12}$
and $\sin\theta_{13}$)   with the CP violating phase $\delta$
in Figure-1. From the figure, it can be seen that in this formalism,  it is possible to accommodate simultaneously
the observed value of the reactor mixing angle
$\theta_{13}$ and solar mixing angle $\theta_{12}$.
The correlation plots between the solar and atmospheric mixing angles with $\theta_{13}$ is shown in Figure-2. In Figure-3, we
show the variation of the Jarlskog Invariant  $J_{CP}$ with $\delta$ and $\theta_{13}$. From the figure it can be seen that
it could be possible to have large CP violation ${\cal O}(10^{-2})$ in the lepton sector.
The correlation between the Dirac CP violating phase
$\delta_{CP}$ and the CP violating parameter $\delta$ of the charged lepton mixing matrix is shown in Figure-4. 
The variation of $M_{ee}$ with the lightest neutrino mass $m_1$
(for Normal Hierarchy) and the variation of $M_{\beta}$ with $\sum m_i$ (where the parameters are varied in their $1~\sigma$ range) are shown 
in Figure-5. Thus, for $m_1$ below ${\cal O}(10^{-2})$ eV, one can get $M_{ee} \leq 1.2 \times 10^{-2}$ eV and $M_\beta \leq 1.4 \times 10^{-2}$ eV.

\begin{figure}[htb]
\includegraphics[width=8cm,height=6cm, clip]{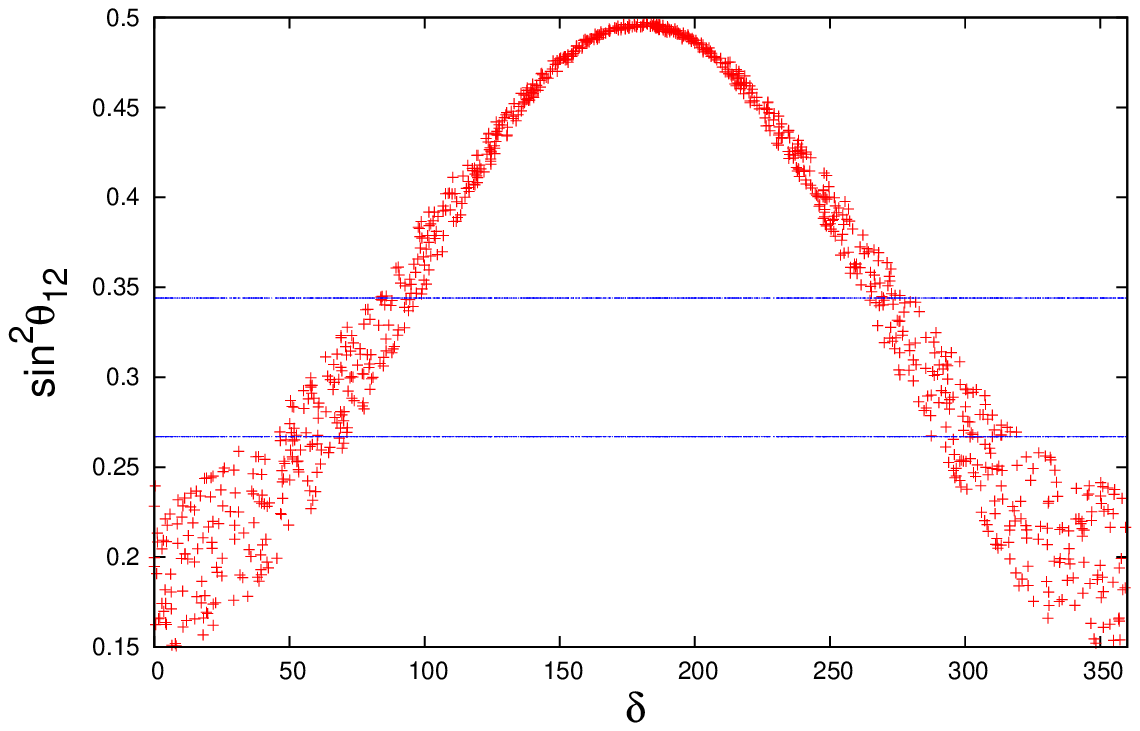}
\hspace{0.2 cm}
\includegraphics[width=8cm,height=6cm, clip]{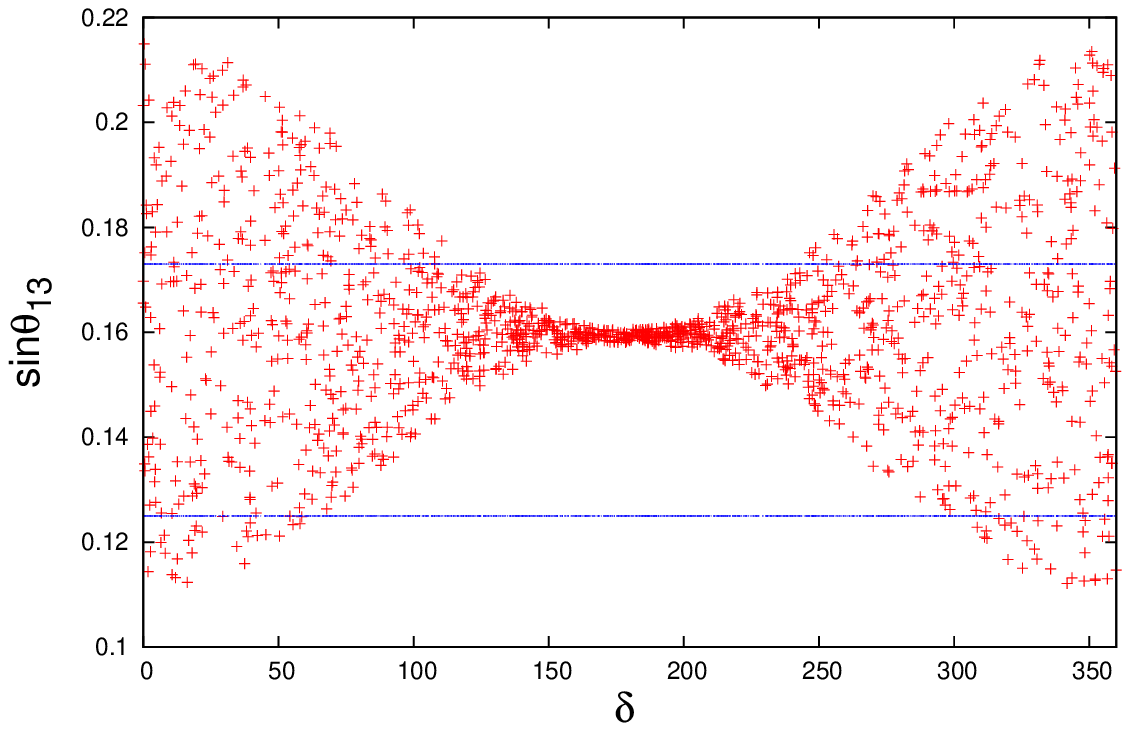}
\caption{Variation of $\sin^2 \theta_{12}$  with
the  CP violating phase $\delta$ (left panel) and $\sin \theta_{13}$ on the right panel. The horizontal lines
(in both panels) represent the $3~\sigma $ allowed range.}
\end{figure}

\begin{figure}[htb]
\includegraphics[width=8cm,height=6cm, clip]{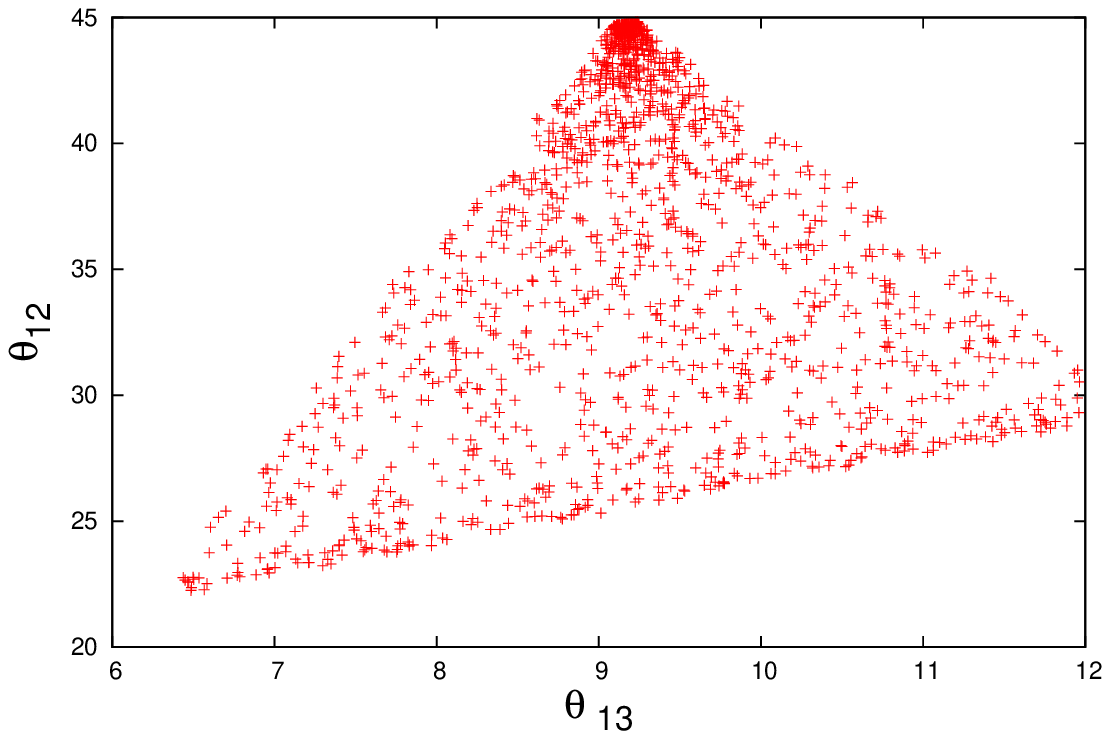}
\hspace{0.2 cm}
\includegraphics[width=8cm,height=6cm, clip]{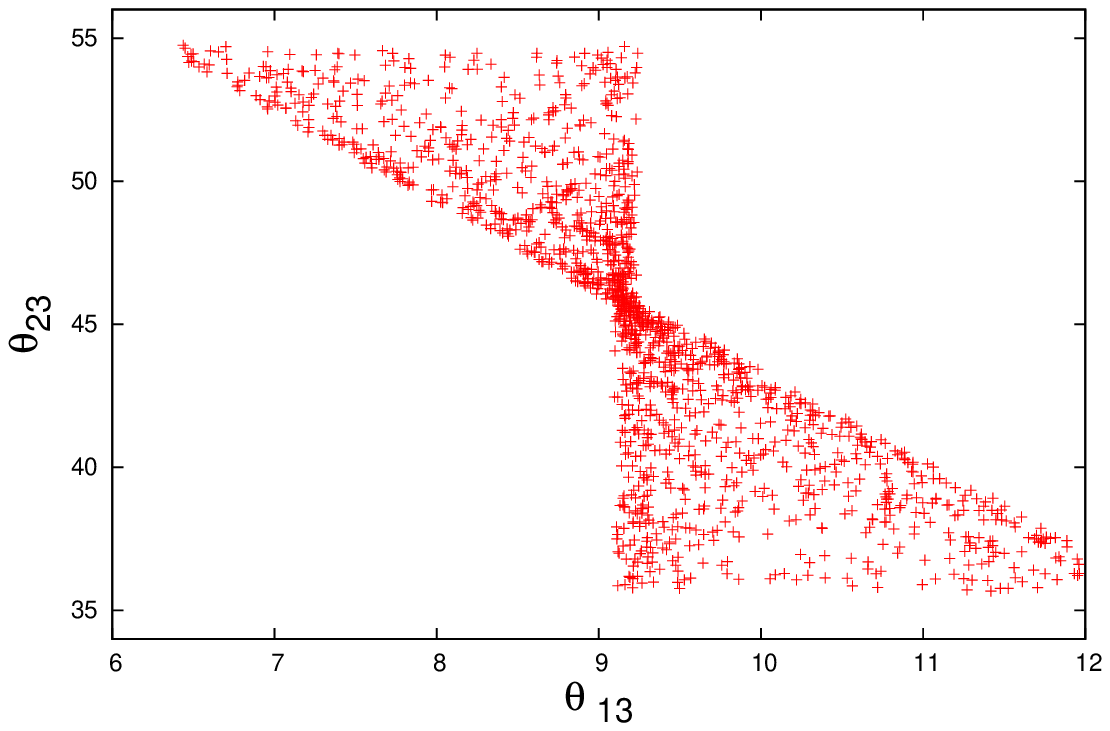}
\caption{Correlation plot between solar (left panel) and the atmospheric mixing angle (right panel)
with $\theta_{13}$. }
\end{figure}

\begin{figure}[htb]
\includegraphics[width=8cm,height=6cm, clip]{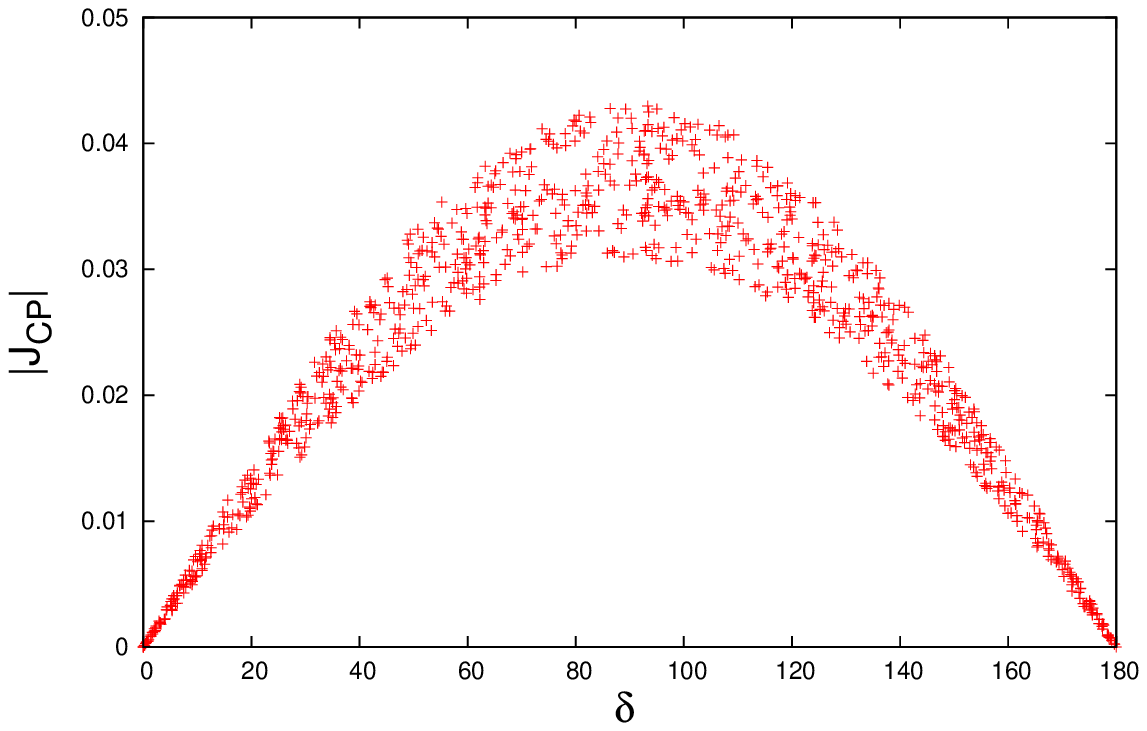}
\hspace{0.2 cm}
\includegraphics[width=8cm,height=6cm, clip]{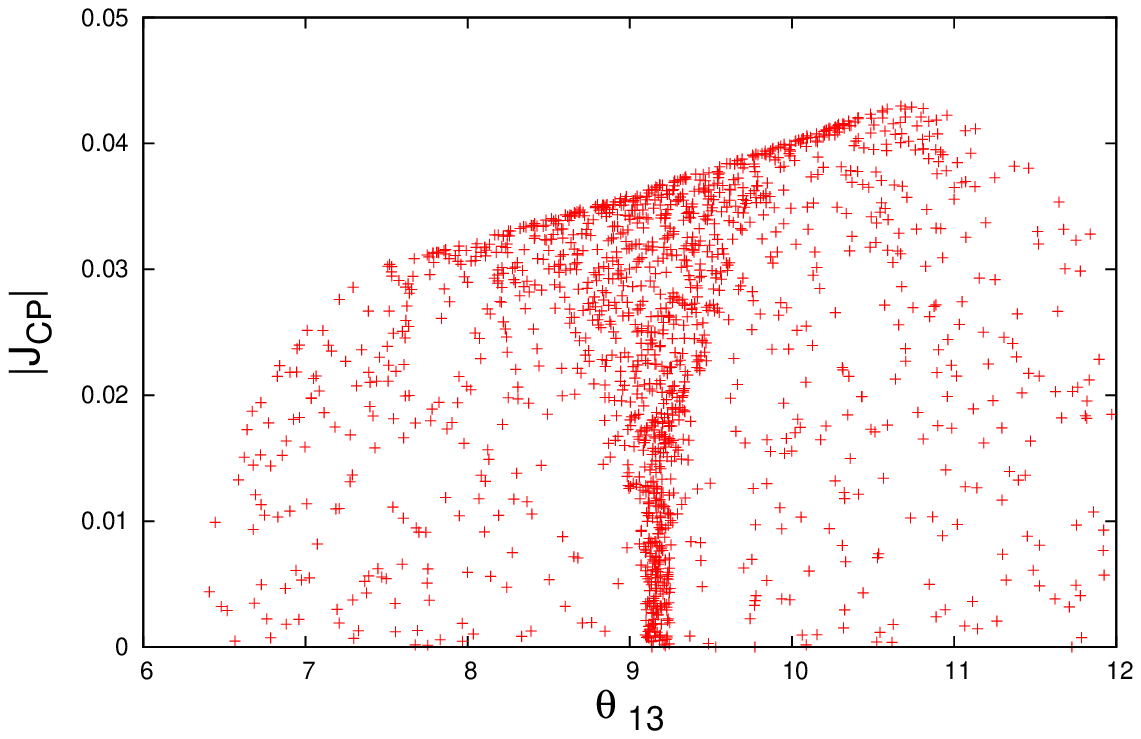}
\caption{Variation of $J_{CP}$ with $\delta$ (left panel) and with
$\theta_{13}$  (right panel). }
\end{figure}

\begin{figure}[htb]
   \centerline{\epsfysize 2.5 truein \epsfbox{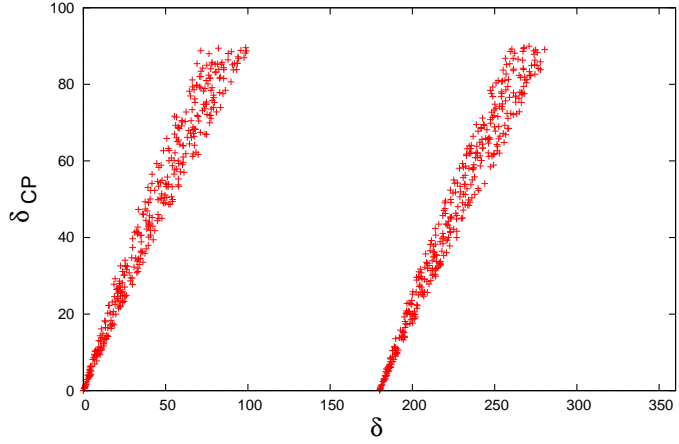}}
 \caption{
  The correlation plot between the Dirac CP violating phase $\delta_{CP}$ and $\delta$.}
  \end{figure}
\begin{figure}[htb]
\includegraphics[width=8cm,height=6cm, clip]{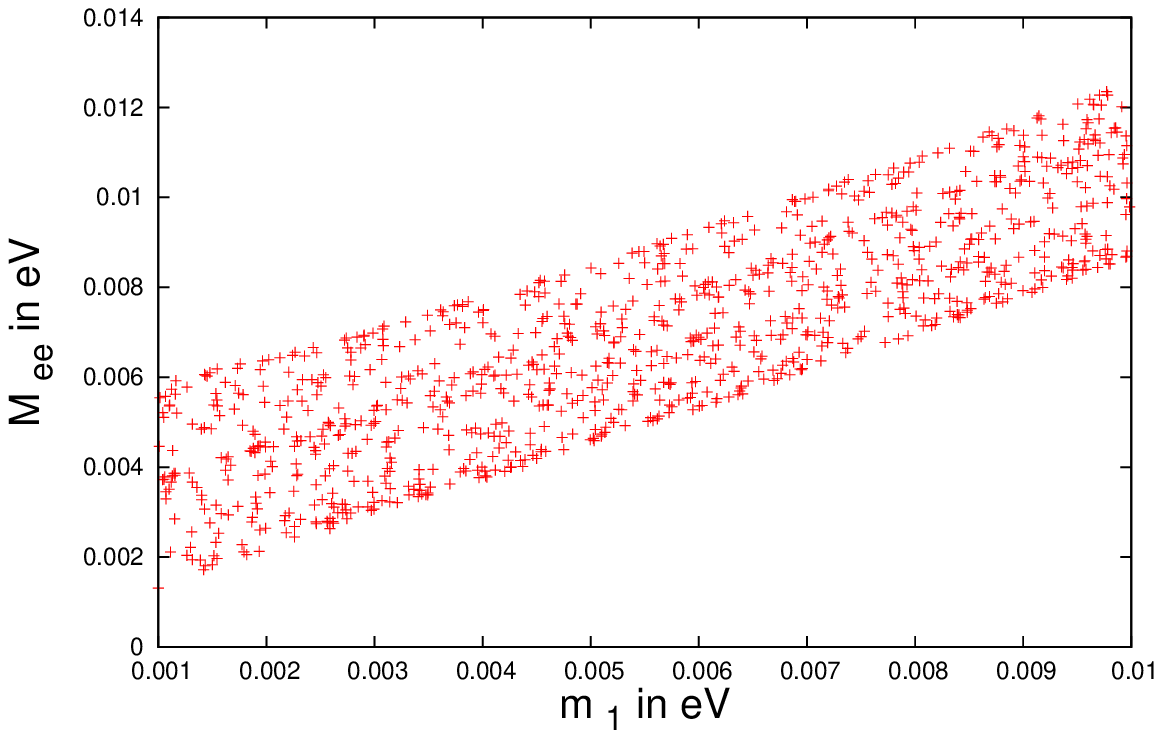}
\hspace{0.2 cm}
\includegraphics[width=8cm,height=6cm, clip]{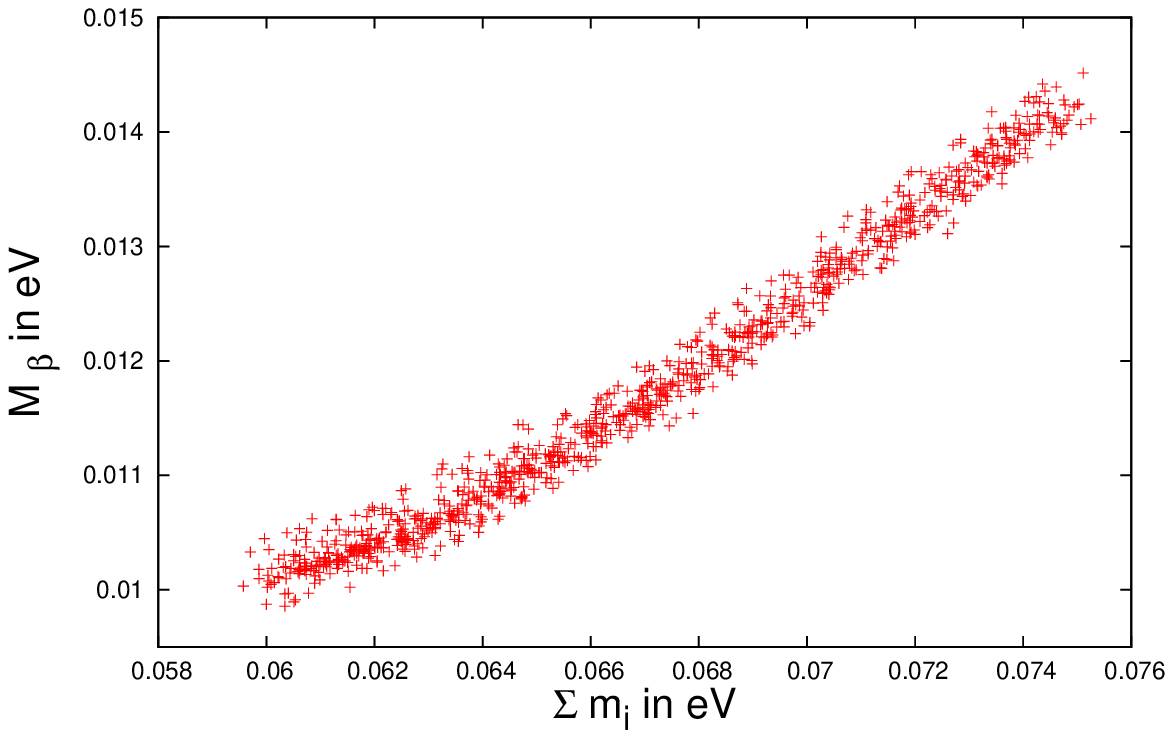}
\caption{Variation of  $M_{ee}$ with the lightest neutrino mass $m_1$
(left panel) and the variation of $M_{\beta}$ with $\sum m_i$ (right panel).}
\end{figure}

To summarize, to accommodate the observed value of relatively large $\theta_{13}$, we consider the corrections
due to the charged lepton mixing matrix to the TBM pattern of neutrino mixing matrix.
Based on the possible inter-relation between the charged lepton and the quark mixing structures
we constructed the lepton mixing matrix to have the form of the CKM-like matrix
(induced from the charged lepton sector) times the TBM  matrix induced from the
neutrino sector. Our result showed that in this formalism, it is possible to accommodate the observed reactor mixing angle $\theta_{13}$
along with the other mixing parameters within their experimental range.
 We have also found that sizable leptonic CP violation characterized by the Jarlskog invariant $J_{CP}$, i.e., $|J_{CP}| \leq
10^{-2}$ could be possible in this scenario. 
The observation of CP violation in the upcoming long base-line neutrino experiments would be a
smoking gun signal of this formalism.
We have also shown that the measured value of $\theta_{13}$ along with
other mixing parameters can be used for constraining the value
of the Dirac CP violating phase $\delta_{\rm CP}$.
The upper limits on $M_{ee}$ and $M_\beta$ are found to be ${\cal O}(10^{-2})$, if the mass of the lightest neutrino
$m_1 \leq 0.01$ eV.

{\bf Acknowledgments}
KND  would like to thank University Grants Commission for financial support.
The work of RM was partly supported by the Council of Scientific and Industrial Research,
Government of India through grant No. 03(1190)/11/EMR-II.

\end{document}